\documentclass[final]{aipproc}\layoutstyle{6x9}
\usepackage{amssymb, amsmath, amsfonts,epsfig,graphics}

\usepackage{bm}
\begin{document}

\def\bsig{\mbox{\boldmath $\sigma$}}
\def\bsig{\mbox{\boldmath $\Sigma$}}
\def\bgam{\mbox{\boldmath $\gamma$}}
\def\bgam{\mbox{\boldmath $\Gamma$}}
\def\bphi{\mbox{\boldmath $\phi$}}
\def\bphi{\mbox{\boldmath $\Phi$}}
\def\btau{\mbox{\boldmath $\tau$}}
\def\btau{\mbox{\boldmath $\Tau$}}
\def\btau{\mbox{\boldmath $\partial$}}
\def\Delc{{\Delta}_{\circ}}
\def\bp{\mid {\bf p} \mid}
\def\al{\alpha}
\def\bet{\beta}
\def\gam{\gamma}
\def\del{\delta}
\def\Del{\Delta}
\def\te{\theta}
\def\nua{{\nu}_{\alpha}}
\def\nui{{\nu}_i}
\def\nuj{{\nu}_j}
\def\nue{{\nu}_e}
\def\num{{\nu}_{\mu}}
\def\nut{{\nu}_{\tau}}
\def\2te{2{\theta}}
\def\chic#1{{\scriptscriptstyle #1}}
\def\chicl{{\chic L}}
\def\lam{\lambda}
\def\SU{SU(2)_{\chic L} \otimes U(1)_{\chic Y}}
\def\Lam{\Lambda}
\def\sig{\sigma}
\def\O{\Omega}
\def\o{\omega}
\def\s{\sigma}
\def\D{\Delta}
\def\d{\delta}
\def\df{\rm d}
\def\8{\infty}
\def\ld{\lambda}
\def\eps{\epsilon}

\newcommand{\be}{\begin{equation}}
\newcommand{\ee}{\end{equation}}
\newcommand{\ba}{\begin{array}}
\newcommand{\ea}{\end{array}}
\newcommand{\dis}{\displaystyle}
\newcommand{\alfad}{\frac{\dis \bar \alpha_s}{\dis \pi}}
\newcommand{\bra}{\mbox{$<$}}
\newcommand{\ket}{\mbox{$>$}}

\title{Rare top quark decays in extended models}

\author{R. Gait\'an}
{address={Centro de Investigaciones Te\'oricas,\\
                 Facultad de Estudios Superiores -- Cuautitl\'an,\\
                 Universidad Nacional Aut\'onoma de M\'exico, (FESC-UNAM).\\
         A. Postal 142,Cuautitl\'an-Izcalli, Estado de M\'exico, 
C. P. 54700, M\'exico.}}
\author{O. G. Miranda}
{address= {Departamento de F{\'\i}sica,\\
                 Centro de Investigaci\'on y de Estudios Avanzados del IPN\\
                 A. Postal 14-740, M\'exico D. F. 07000, M\'exico.}}
\author{L. G. Cabral-Rosetti}
{address = {
Departamento de Posgrado,\\ 
Centro Interdisciplinario 
de Investigaci\'on y Docencia en Educaci\'on T\'ecnica (CIIDET),\\
Av. Universidad 282 Pte., Col. Centro, A. Postal 752,\\
C. P. 76000, Santiago de Queretaro, Qro., M\'exico.}}
\begin{abstract}
Flavor changing neutral currents (FCNC) decays $t \rightarrow H^0
+ c$, $t \rightarrow Z + c$, and $H^0 \rightarrow t + {\bar c}$ are
discussed in the context of Alternative Left-Right symmetric Models
(ALRM) with extra isosinglet heavy fermions where FCNC decays may take
place at tree-level and are only suppressed by the mixing between
ordinary top and charm quarks, which is poorly constraint by current
experimental values.  The non-manifest case is also briefly discussed.
\end{abstract}
\pacs{14.65.Ha,14.80.Cp,12.60.Cn,12.15.Ff}

\maketitle

\section{Introduction}

Flavor-changing neutral currents (FCNC) are absent in the Standard
Model (SM) at the tree-level due to the Glashow-Iliopoulos-Maiani
(GIM) mechanism. However, new FCNC states can appear in top decays if
there is physics beyond the Standard Model. In this context, rare top
quark decays are interesting because they might be a source of
possible new physics effects.  In some particular models beyond the
SM, rare top decays may be significantly enhanced to reach detectable
levels \cite{Bejar1}.

Rare top decays have been studied in the context of the SM and beyond
\cite{Diaz:91,Jenkins,Mele}. The top quark decays into gauge bosons
($t \rightarrow c + V;\, V \equiv \gamma,\, Z,\, g$) are extremely
rare events in the SM. 
However, by considering physics beyond the SM,
for example, the Minimal Supersymmetric Standard Model (MSSM) or the
two-Higgs-doublet model (2HDM) or extra quark singlets, new
possibilities open up \cite{Bejar1, Aguilar-Saavedra:2002kr},
enhancing this branching ratios to the order of $\sim 10^{-6}$ for the
$t\to c + Z$~\cite{Aguilar} channel and $\sim10^{-4}$ for the $t\to c+
H$~\cite{Herrero} case. 

In the future CERN Large Hadron Collider (LHC), about $10^7$ top quark
pairs will be produced per year~\cite{Beneke}. An eventual signal of
FCNC in the top quark decay will have to be ascribed to new
physics. Furthermore, since the Higgs boson could also be produced at
significant rates in future colliders, it is also important to search
for all the relevant FCNC Higgs decays.

On the other hand, while the electroweak SM has been successful in the
description of low-energy phenomena, it leaves many questions
unanswered. One of them has to do with the understanding of the origin
of parity violation in low-energy weak interaction processes. Within
the framework of left-right symmetric models, based on the gauge group
$SU(2)_L \otimes SU(2)_R \otimes U(1)_{B-L}$, this problem finds a
natural answer\cite{pati-salam,Mohapatra-book}. Moreover, new
formulations of this model have been considered in which the fermion
sector has been enlarged to include isosinglet vectorlike heavy
fermions in order to explain the mass
hierarchy~\cite{Davidson-Wali,Kiers}. 
Most of these models includes two Higgs doublets.

We consider the rare top decay into a Higgs boson and
the FCNC decay of the Higgs boson with the presence of a top quark in
the final state, within the context of these alternative left-right
models (ALRM) with extra isosinglet heavy fermions. Due to the
presence of extra quarks the Cabibbo-Kobayashi-Maskawa matrix is not
unitary and FCNC may exist at tree-level.

\section{The Model}

The ALRM formulation is based on the gauge group $SU(2)_{\chic L}
\otimes SU(2)_{\chic R} \otimes U(1)_{B-L}$. In order to solve
different problems such as the hierarchy of quark and lepton masses or
the strong CP problem, different authors have enlarged the fermion
content to be of the form

\begin{eqnarray}
l^{\chic {0}}_{i\, {\chic {L}}} =
     \left( \begin{array}{c}
     \nu^{\chic {0}}_i \\ {e^{\chic {0}}_i}\end{array} \right)_{\chic {L}}
   ,\ {e}^{\chic {0}}_{i\,{\chic {R}}}\ \ \ \ \ &;& \ \ \ \ \
 {\widehat l}^{\chic {0}}_{i\, {\chic {R}}} =
     \left( \begin{array}{c}
     {\widehat {\nu}^{\chic {0}}_{i}} \\
     {{\widehat {e}}^{\chic {0}}_{i}} \end{array} \right)_{\chic {R}}
   ,\ {\widehat {e}}^{\chic {0}}_{i\,{\chic {L}}}  \nonumber \\
Q^{\chic {0}}_{i\, {\chic {L}}} =
     \left( \begin{array}{c}
     {{u}^{\chic {0}}_{i}}\\
     {{d}^{\chic {0}}_{i}}
     \end{array} \right)_{\chic {L}}
   ,\ {u}^{\chic {0}}_{i\,{\chic {R}}}\
   ,\ {d}^{\chic {0}}_{i\,{\chic {R}}}, \ \ \ \ \ &;& \ \ \ \ \
{\widehat Q}^{\chic {0}}_{i\, {\chic {R}}} =
     \left( \begin{array}{c}
          {\widehat {u}^{\chic {0}}_{i}}\\
     {{\widehat {d}}^{\chic {0}}_{i}}
     \end{array} \right)_{\chic {R}}
   ,\ {\widehat {u}}^{\chic {0}}_{i\,{\chic {L}}}\
   ,\ {\widehat {d}}^{\chic {0}}_{i\,{\chic {L}}}\ ,
\label{Eq2.1}
\end{eqnarray}

where the index $i$ ranges over the three fermion families.
The superscript $0$ denote weak eigenstates. 
In many of these models, extra neutral leptons also appears in order
to explain the neutrino mass pattern, however we will focus in this
work only on the quark sector.

In order to break $SU(2)_{\chic L} \otimes SU(2)_{\chic R} \otimes
  U(1)_{B-L}$ down to $U(1)_{\chic {em}}$ the ALRM introduces
  two Higgs doublets, the SM one ($\phi$) and its partner (${\widehat
  \phi}$). 
Ref. \cite{Ceron} shows that from the eight scalar degrees of freedom,
six become the Goldstone bosons required to give mass to the $W^\pm$,
${\widehat W}^\pm$, $Z$ and ${\widehat Z}$; thus two neutral Higgs
bosons, $H$ and $\widehat{H}$, remain in the physical spectrum. 

The renormalizable and gauge invariant interactions of the scalar
doublets $\phi$ and ${\widehat \phi}$ with the fermions are described
by the Yukawa Lagrangian. For the quark fields, the corresponding
Yukawa terms are written as

\be
\ba{c}
\displaystyle
{\cal L}_{\chic {Y}}^{q} = {\lam}^{d}_{i j}\,
{\overline {Q^{\, {\chic 0}}_{i {\chic L}}}}\,
{\phi}\, d^{\, {\chic 0}}_{j {\chic R}}
+ {\lam}^{u}_{i j}\, {\overline {Q^{\, {\chic 0}}_{i {\chic L}}}}\,
{\widetilde {\phi}}\, u^{\, {\chic 0}}_{j {\chic R}}
+ {\widehat \lam}^{d}_{i j}\,
{\overline {\widehat Q^{\, {\chic 0}}_{i {\chic R}}}}\,
{\widehat \phi}\, {\widehat d}^{\, {\chic 0}}_{j {\chic L}}
\\[0.3cm]
\displaystyle
+ {\widehat \lam}^{u}_{i j}\,
{\overline {\widehat Q^{\, {\chic 0}}_{i {\chic R}}}}\,
{\widetilde {\widehat {\phi}}}\, {\widehat u}^{\, {\chic 0}}_{j {\chic L}}
+ {\mu}^{d}_{i j}\, {\overline {\widehat d^{\, {\chic 0}}_{i {\chic L}}}}\,
d^{\, {\chic 0}}_{j {\chic R}}
+ {\mu}^{u}_{i j}\, {\overline {\widehat u^{\, {\chic 0}}_{i {\chic L}}}}\,
u^{\, {\chic 0}}_{j {\chic R}} + h. c.
\ea
\label{Eq2.5}
\ee where $i,j = 1, 2, 3$ and ${\lam}^{d (u)}_{i j}$, ${\widehat
  \lam}^{d (u)}_{i j}$, and ${\mu}^{d (u)}_{i j}$ are (unknown)
matrices. The conjugate fields ${\widetilde {\phi}}$ $\Big(
{\widetilde {\widehat {\phi}}} \Big)$ are ${\widetilde {\phi}} = i
\tau_2 \phi^*$ and ${\widetilde {\widehat {\phi}}} = i \tau_2
{\widehat \phi^*}$, with $\tau_2$ the Pauli matrix.

We can introduce the generic vectors $\psi^0_L$ and
$\psi^0_R$~\cite{Langacker:1988ur} , for representing left and right
electroweak states with the same charge. These vectors can be
decomposed into the ordinary $\psi^0_{OL}$ and the exotic
$\psi^0_{EL}$ sectors

\be
\psi^0_L =
\left(\begin{array}{c}
\psi^0_{OL} \\ \\ \psi^0_{EL}  \end{array}\right)
\qquad \psi^0_R =
\left(\begin{array}{c}
\psi^0_{OR} \\ \\ \psi^0_{ER}  \end{array}\right),
\ee

In the same way we can define the vectors for the mass eigenstates in
terms of 'light' $\psi_{lL}$ and 'heavy' $\psi_{hL}$ states. 
The relation between weak eigenstates and mass eigenstates will be
given through the matrices $U_L$ and $U_R$ by 
$
\psi^0_L = U_L \psi_L$, $\psi^0_R =U_R \psi_R
$
where

\be
{\sf U}_a = \left(\begin{array}{ccc}
{{\sf A}_a} &  & {{\sf E}_a}  \\
{{\sf F}_a} &  & {{\sf G}_a}
\end{array}\right)
, \qquad a = L,R
\label{Eq2.7}
\ee
Here, $A_a$ is the $3\times 3$ matrix relating the ordinary weak
states with the light-mass eigenstates, $G_a$ is a $3\times3$ matrix
relating the exotic states with the heavy ones, while $E_a$ and $F_a$
describe the mixing between the two sectors.

In this model, thanks to the extra heavy quarks,  it is
possible to have a relatively big mixing between ordinary quarks. This
is not a particular characteristic of the model but a general feature when
considering models with extra heavy singlets~\cite{Aguilar-Saavedra:2004wm}.

The tree-level interaction of the neutral Higgs bosons $H$ and
${\widehat H}$ with the light fermions are given by

\be
\ba{c}
\displaystyle
{\cal L}_{\chic {Y}}^{f} = \frac{g}{2 \sqrt 2}
{\overline \psi_{\chic L}} A^{\dag}_{\chic L} A_{\chic L}
\frac{m_f}{M_{\chic W}} \psi_{\chic R} \Big( H\, \cos \alpha
- {\widehat H}\, \sin \alpha \Big)
\\[0.3cm]
+ \displaystyle
\frac{{\widehat g}}{\sqrt 2}
{\overline \psi_{\chic L}} \frac{m_f}{M_{\chic {\widehat W}}}
F^{\dag}_{\chic R} F_{\chic R} \psi_{\chic R}\Big( H\, \sin \alpha
+ {\widehat H}\, \cos \alpha \Big) + h. c.
\ea
\label{Eq2.6}
\ee

The neutral current in terms of the mass eigenstates, including the
contribution of the neutral gauge boson mixing, can be written directly 
from this Lagrangian. 

From the last equation we can see that, thanks to the
non-unitarity of the $A_{a}$ matrices we can have FCNC at tree-level.
This characteristic appears due to the extra quark content
of the model, which is not present in the usual left-right symmetric
model.

\section{FCNC top and Higgs decays in the ALRM}

Once we have introduced the model in which we are interested, we
compute the expected branching ratio for a FCNC top or Higgs decay
with a charm quark in the final state. We perform this analysis in
this section.

\subsection{Constraining the top-charm mixing angle}

In order to have an expectation on the branching ratio for the FCNC
top decay in the ALRM we need first an estimate on the mixing between
the top and charm quarks in the model. One may think that the best
constrain could come from the flavor-changing coupling of the
neutral $Z$ boson to the top and charm quarks, which can be written
as:

\be
\ba{c}
\displaystyle
{\cal L}_{Z}^{c t} = \frac{e}{s_{\theta_{\chic W}}\, c_{\theta_{\chic W}}}\,
\overline c\, (g_{\chic V} + g_{\chic A}\gamma^5)\, \gamma^{\mu}\, Z_{\mu}\, t
\ea
\label{eq:Lag.tzc}
\ee
where

\be \ba{c} g_{\chic {V}, {\chic A}} = \frac{1}{4}\, (c_{\Theta} -
\frac{g}{\widehat {g}} \frac{s^{2}_{\theta_{\chic
W}}}{r_{\theta_{\chic W}}}\, s_{\Theta})\, \eta^{L}_{\chic {32}}
\pm \frac{1}{4}\, \frac{\widehat {g}}{g}
\frac{c^{2}_{\theta_{\chic W}}}{r_{\theta_{\chic W}}}\,
s_{\Theta}\, \eta^{R}_{\chic {32}} \ea
\label{gvtc} \ee
and $s_{\theta_{\chic W}}$, $c_{\theta_{\chic W}}$ and
$r_{\theta_{\chic W}}$ are, respectively, $\sin\theta_{\chic W}$,
$\cos\theta_{\chic W}$ and $\sqrt{\cos^{2}{\theta_{\chic W}} -
(g^2/{\widehat{g}}^2)\sin^{2}{\theta_{\chic W}}}$; $\theta_{\chic W}$
is the weak mixing angle, $\Theta$ is the mixing between the $Z$ and
$\widehat{Z}$ neutral gauge bosons.
Here, $\eta^L_{\chic {32}}$ and $\eta^{R}_{\chic {32}}$ represent the mixing
between the ordinary top and charm quarks and are given by
\be \eta^L_{\chic {32}} = (A^{+}_{L}\, A_{L})_{\chic
    {32}}\qquad
\eta^R_{\chic {32}}
= (A^{+}_{R}\, A_{R})_{\chic
    {32}}.  \ee
The mixing between the $Z$ and the $\widehat{Z}$ neutral gauge bosons,
$\Theta$, is expected to be small~\cite{Adriani} if the ratio $r_g =
g/{\widehat{g}} = 1$~. However, one might think that for different
values of $r_g$ these bounds are not longer valid. This is actually
true, however, for most of the values of $r_g$, the expected freedom
for the mixing angle $\Theta$ is still limited.

We can see from the definition of $r_{\theta_{\chic W}}$ that the
value of this ratio can not be bigger than
$\sqrt{\cos^{2}{\theta_{\chic W}} / \sin^{2}{\theta_{\chic W}}}\approx
1.82$. We can recalculate the constraint obtained in~\cite{Adriani}
taking into account the freedom of this parameter.  In order to do
this analysis we simply need to consider the appropiate range for the
parameter $r_g$ that will affect the coupling constants for the $Z\to
e^+e^-$ that are needed for such computation: 
\be \ba{c} g^e_{\chic {V}, {\chic A}} = -\frac{1}{2}\, (c_{\Theta} -
\frac{g}{\widehat {g}} \frac{s^{2}_{\theta_{\chic
W}}}{r_{\theta_{\chic W}}}\, s_{\Theta})
\pm \frac{1}{2}\, \frac{\widehat {g}}{g}
\frac{c^{2}_{\theta_{\chic W}}}{r_{\theta_{\chic W}}}\,
s_{\Theta}  \ea
\label{gvee} . \ee 
Note that in this case we are not taking into account the lepton
flavor violation that has been discussed in ref~\cite{Ceron}.

\begin{figure*}
\includegraphics[width=0.5\textwidth,angle=-90]{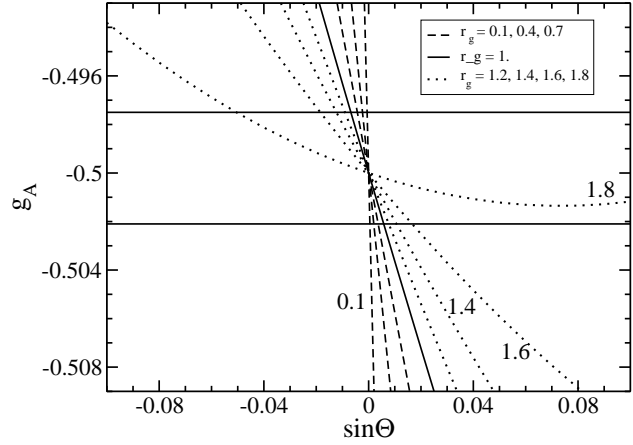}
       \caption{
Expected value of $g_A$ for the model in dependence on the mixing angle 
between $Z$ and  
$\widehat{Z}$, and the ratio $r_g= g/\widehat{g}$. The horizontal lines 
shows the 90 \% C. L. allowed by the experiment.
    \label{fig:rg}}
\end{figure*}

With this formula and the limit for $g_A$ obtained from the
experiment: $g_A^{\rm exp}= -0.4998\pm 0.00014$~\cite{Acciarri} we can
obtain a constraint for the $\Theta$ mixing depending on the value of
$r_g$. The result of such analysis is shown in Fig. (\ref{fig:rg})
were it is possible to see that $|\sin\Theta|\leq 0.03$, if the value
of the ratio $r_g$ is smaller than $1.6$. As $r_g$ approaches the
critical value of $1.8$, the constraint will disappear.

Therefore, the mixing angle $\Theta$ can be safely neglected for most
of the values of $g$ and $\widehat{g}$. In this case the expression in
Eq. (\ref{gvtc}) will not depend on the parameter $\eta^R_{\chic
{32}}$. For the present analysis we will consider only the case with
$g = \widehat{g}$, and therefore from now on we will denote
$\eta_{\chic {32}} = \eta^L_{\chic {32}}$.

From Eq. (\ref{eq:Lag.tzc}) we can compute the branching ratio for the
decay $t \to Z +c$ and compare it to the experimental limit $B(t
\rightarrow Z + c) \leq 0.137$~\cite{Abbiendi} at 95 \% C. L. We will
get the maximum value for $\eta_{\chic {32}} \leq 0.53$.

Although we have found a direct constrain to $\eta_{\chic {32}}$, it
is possible to get a stronger limit if we use the unitarity properties
of the mixing matrix and the constrain on $\eta_{\chic {22}}$ that
comes from the branching ratio $\Gamma (Z \rightarrow c + \bar{c})$.
The experimental value for the branching ratio of this process is
given by $B(Z\to c\bar{c})=\Gamma(Z\to c\bar{c})/\Gamma_{\chic{
total}}=0.1181 \pm 0.0033$ (see~\cite{pdg}). Using this experimental
value, the minimum value for $\eta_{\chic {22}}$ at 95 \% C. L. will be
$\eta_{\chic {22}} \ge 0.99$.

This information is of great help for constraining $\eta_{\chic{32}}$
since the unitarity of the mixing matrix has already been analyzed in
the general case~\cite{Aguila-PRL} and leads to the relation
$
|\eta_{\chic{32}}|^2 \le (1 - \eta_{\chic{33}})(1 - \eta_{\chic{22}}).
\label{eq:unitary}
$

Although we don't know the value for $\eta_{\chic{33}}$, the boundary
on $\eta_{\chic{22}}$ is enough to see that the mixing parameter
$\eta_{\chic{32}} \le 0.1$. The higher value $\eta_{\chic{23}} = 0.1$
is obtained when we take the extreme case $\eta_{\chic{33}} = 0$, as
can be seen from the equation in the previous paragraph. 

It is possible to obtain more stringent constraints if low-energy data
are considered. For the case of two extra quark singlets, this
analysis was done in a very general framework in
Ref. \cite{Aguilar-Saavedra:2002kr}. After a very complete analysis of
all the observables, the author of this article obtained
$|\eta_{\chic{32}}|\leq 0.036$. This relatively large value is allowed
for the case of a exotic top mass similar to that of the SM
top-quark (There are not stringent lower bounds on the mass
of a exotic top quark, being $220$~GeV the current direct
limit~\cite{Acosta03}). In the case of a very heavy mass for the
exotic top-quark the constraint is more stringent:
$|\eta_{\chic{32}}|\leq 0.009$. In what follows we will use these two
values in order to illustrate the expected signals from rare Higgs and
top decays.

\subsection{The decays $t \rightarrow H^0 + c$ and 
$H^0 \rightarrow t + \bar{c}$}

Now that we have an estimate for the value of $\eta_{\chic {32}}$, we
compute the branching ratio for $t \rightarrow H^0 + c$ in the
framework of ALRM. We take the charged-current two-body decay $t
\rightarrow b + W$ to be the dominant t-quark decay mode. The neutral
Higgs boson $H^0$ will be assumed to be the lightest neutral mass
eigenstate.

Assuming $M_{\widehat{H}} \gg M_H$ the vertex $tcH^{0}$ is written
as 
$
\frac{g\, m_t\, \eta_{\chic {32}}}{2 M_{\chic {W}}}\, \cos \alpha \, P_L.
$
The partial width for this tree-level process can be obtained
in the usual way and it is given by:
\be
\ba{c}
\displaystyle
\displaystyle
  \frac{G_{\chic F}\, \eta^2_{\chic {32}}\, \cos^2
\alpha} {16\, \sqrt{2}\, \pi\, m_t}\, \Big( m_t^2 +  m_c^2 -
M_{\chic H}^2 \Big ) \Big[ \Big( m_t^2 -  \Big( M_{\chic H} + m_c
\Big )^2\, \Big) \Big( m_t^2 -  \Big( M_{\chic H} - m_c  \Big)^2\,
\Big) \Big]^{\frac{1}{2}}
\ea
\label{ricard}
\ee
where $G_{\chic F}$ is the Fermi's constant, $m_t$ denotes the top
mass, $m_c$ is the charm mass, and $M_{\chic H}$ is the mass of the
neutral Higgs boson.  We can see from this formula that the branching
ratio will be proportional to the product $\eta_{\chic {32}}\cos
\alpha$, of the top-quark mixing with the SM Higgs boson mixing with
the extra Higgs boson.

The branching ratio for this decay is obtained as the ratio of Eq.
(\ref{ricard}) to the total width for the top quark, namely
$
\ba{c}
B(t \rightarrow H^{0} + c)=
\frac{\Gamma(t \rightarrow H^{0} + c)}{\Gamma(t \rightarrow b + W)} .
\ea
$

Thanks to the possible combined effect of a big $\cos\alpha$ (null
mixing between the SM Higgs boson and the additional Higgs bosons) and
a big value of $\eta_{\chic {32}}$ this branching ratio could be as
high as $\approx 3\times 10^{-4}$, for a Higgs mass of
117~$GeV$. Perhaps is more realistic to consider the more stringent
constraint $\eta_{\chic {32}} = 0.009$, but even in this case, for
$\cos \alpha \approx 1$ there is still sensitivity for detecting a
positive signal of order $10^{-5}$.

Finally we also consider the case of a Standard Higgs with a large
mass. The best-fit value of the expected Higgs mass, including the new
average for the mass of the top quark, is 117 GeV~\cite{Abazov:2004cs}
and the upper bound is $M_H \le 251$~GeV at 95 \% C L. However, the
error for the Higgs boson mass from this global fit is asymmetric, and
a Higgs mass of $400$~GeV is well inside the $3\sigma$ region as can
be seen in Ref~\cite{Abazov:2004cs}.

We estimate the branching ratio for the decay $H^0 \rightarrow t +
\bar{c}$, where $H^{0}$ is the light neutral Higgs boson of the
ALRM. The expression for the partial width is

\be
\ba{c}
\displaystyle
\displaystyle
 \frac{3\, G_{\chic F}\, m_t^2\, \eta^2_{\chic
{32}}\, \cos^2 \alpha} {8\, \sqrt{2}\, \pi\, M_{\chic H}^3}\,
\Big( M_{\chic H}^2 - m_t^2 -  m_c^2 \Big ) \Big[ \Big( M_{\chic H}^2
-  \Big( m_t + m_c \Big )^2\, \Big) \Big( M_{\chic H}^2 -
\Big( m_c - m_t \Big )^2\, \Big) \Big]^{\frac{1}{2}}.
\ea
\label{Eq2.11}
\ee

The branching ratio for this decay is obtained as the ratio of Eq.
(\ref{Eq2.11}) to the total width of the Higgs boson, which will
include the dominant modes $H^{0} \rightarrow b + \bar{b}$, $H^{0}
\rightarrow c + \bar{c}$, $H^{0} \rightarrow \tau + \bar{\tau}$,
$H^{0} \rightarrow W + W$, and $H^{0} \rightarrow Z + Z$.  The
expressions for these decay widths in the ALRM also includes
corrections due to the new parameters introduced in the model, and
they are taken into account~\cite{Gaitan}.

We computed the branching ratios for different decay modes, both for
the Standard Model case ($\eta_{\chic {32}} = 0$ and $\eta_{ii}=1$)
and for the FCNC case.  We found that, also for a heavy Higgs, there
are chances to either detect or to constrain the mixing angle
parameter $\eta_{\chic {32}}$.  In this case, since all the partial
widths have the same dependence on $\cos^{2} \alpha$, the branching
ratios will depend only on $\eta_{\chic {32}}$.

\section{Results and conclusions}

The ALRM allows relatively big values of
$\eta_{\chic {32}}$. The $t \to H + c$ branching ratio could be of
order of $10^{-4}$, which is at the reach of LHC.
It has been estimated that the LHC sensitivity (at 95 \%
C. L.)  for this decay  is $Br (t\to Hc)\leq4.5\times
10^{-5}$ \cite{Aguilar-Saavedra:2000aj};
this branching ratio would be obtained in this model for a top-charm
mixing $\eta_{\chic {32}} = 0.015$ and a diagonal ordinary top
coupling $\eta_{\chic {22}} \simeq 0.98$.
On the other hand, the FCNC mode $H \to t + \bar{c}$ may reach a
branching ratio of order $10^{-3} $ and can also be a useful
channel to look for signals of physics beyond the SM in the LHC.


\section{acknowledgments}
This work has been
  supported by Conacyt and SNI.

%

%

\end{document}